# Favorable half-Heusler structure of synthesized TiCoSb alloy: a theoretical and experimental study

Pallabi Sardar[1,2], Suman Mahakal[2], Soumyadipta Pal[3,4], Shamima Hussain[5], Vinayak B. Kamble[6], Pintu Singha[6], Diptasikha Das[1]*, and Kartick Malik[2]*

[1]*Department of Physics, ADAMAS University, Kolkata-700 126, West Bengal, India*
[2]*Department of Physics, Vidyasagar Metropolitan College, Kolkata-700 006, West Bengal, India*
[4]*University of Engineering and Management, New Town, University Area, Plot No. III, B/5, New Town Rd, Action Area III, Newtown, Kolkata 700160, West Bengal, India*
[4]*Department of Physics, Institute of Engineering & Management, Management House, D-1, Sector-V, Saltlake Electronics Complex, Kolkata 700 091, West Bengal, India*
[5]*UGC-DAE Consortium for Scientific Research, Kalpakkam Node, Kokilamedu-603 104, Tamil Nadu, India*
[6]*Indian Institute of Science Education and Research (IISER) Trivandrum, Vithura, Trivandrum 695 551*

**ABSTRACT**

The most favorable structure of the synthesized TiCoSb half-Heusler (HH) alloy is explored theoretically and experimentally and the best structure for thermoelectric conversion is reported. TiCoSb HH alloy is synthesized by solid state reaction method. Rietveld refinement of the X-ray diffraction data employing four probable structures of the HH alloy are performed to obtain the best fitting and identify the crystallized structure. However, microstructural characterization is performed using the energy dispersive X-ray spectroscopy (EDS) and transmission electron microscopy (TEM) to reveal the stoichiometry and Bragg's reflection planes of the synthesized polycrystalline lattice structure of TiCoSb HH alloy. Theoretical investigation is performed by implementing the first principle calculation using the Full Potential Linearized Augmented Plane Wave method (FP-LAPW) in the Quantum Espresso software package. The most probable structure is explored by estimating the minimum energy at equilibrium volume and electronic structure of the TiCoSb HH alloy of the four probable structures, considered. The theoretical and experimental data are corroborated and the most probable structure is identified for the crystallized TiCoSb HH alloy. The thermoelectric properties of the most probable structure are estimated.

## I. Introduction

TiCoSb, a half-Heusler material, is intensively investigated as a potential thermoelectric (TE) material at mid-temperature (500K-900K).[1-3] TE materials draw attention as a solid-state converter of heat to electricity and vice versa.[4-6] The conversion efficiency of a TE material is designated by the term, Figure of Merit, $ZT = \frac{S^2\sigma}{\kappa_L+\kappa_e}$, where S, σ, T, $\kappa_L$ and $\kappa_e$ are the Seebeck coefficient, electrical conductivity, absolute temperature, lattice-thermal conductivity and electronic contribution of thermal conductivity, respectively. The term $S^2\sigma$ is known as the power factor (PF).[6-8] It is crucial to note that optimization of the involved parameters in ZT is required due to their interdependence.[9-11] However, in-depth knowledge on crystal structure,[12-14] electronic band structure,[15] carrier concentration,[16,17] and scattering[18,19] accompanied by electrical conductivity and thermal conductivity is required to improve the PF and ZT.

Various type of TE material viz. chalcogenides,[20-22] skutterudites,[23-25] silicides,[26-28] and clathrates[29,30] are investigated during few decades to obtain the commercially viable TE in the specific temperature ranges. The half-Heusler (HH) alloys [31,32] become a focus of attention as a mid-temperature TE material[33,34] attributing to the narrow band gap,[35] high substitutability of constituent elements,[31] enhanced contact engineering,[36,37] and high thermal stability.[33,38] HH alloys are distinct member of Heusler family, exhibit a wide range of physical properties apart from TE properties. NiMnSb, CoMnSb, and FeMnSb based HH alloys are well known for the magnetic properties and applications as magnetic storage device, spintronics, and magnetic sensors.[39, 40] The efficiency of a photovoltaic device may be enhanced by incorporation of HH alloys as an absorber layer in solar cells or as intermediate band materials.[41, 42] The novel electronic and magnetic properties, arise in HH alloys, are related to vacancies and valence electron count (VEC) in the Heusler structure. The HH compounds may be considered as vacant Wyckoff positions in Heusler alloy with formula X$_2$YZ or XY$_2$Z, corresponding to the structure is presented in figure 1, where X is more electropositive than the transition material Y and the main group material Z.[31, 39, 43] The HH ternary intermetallic with the stoichiometry XYZ consists of three interpenetrating face-centered cubic (fcc) lattice with a vacant position in Heusler alloy.[31, 34] The electronic and transport properties of the HH alloys are strongly correlated with the constituent atoms and structure of the alloy.[31, 44] One of the characteristics of an HH TE material is a narrow band gap semiconductor.[44] Semiconducting properties in 18 VEC HH compounds are manifested due to the exchange of electrons from the most electropositive X to the comparatively electronegative Y and Z elements.[31] The electronegativity of X, Y, and Z atoms are lies in the range 1.2-1.7, 1.8-2.4, and 1.7-2.2, respectively.[45] A close cell configuration, i.e., d$^{10}$ and s$^2$p$^6$, is formed for the Y and Z atoms, resulting in a stable semiconducting HH alloy with VEC 18.[31, 45, 46] However, the stability and semiconducting properties of an HH alloy may be understood by the Zintl chemistry framework, where closed valence cell configuration is achieved by the transfer of valence electron from the electro positive X$^{n+}$ to the tetrahedrally bonded [YZ]$^{n-}$ sublattice.[31]

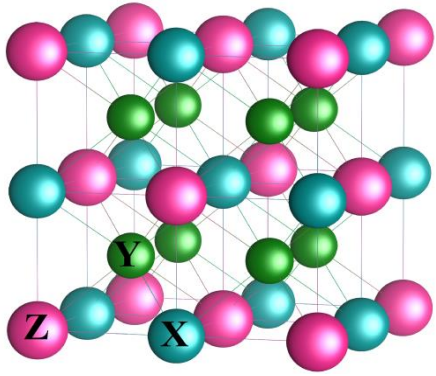

Figure 1: Structure of XY$_2$Z-type Heusler alloy

TiCoSb, a promising HH TE material due to the substantial S and moderate $\sigma$, has been reported experimentally by Gladysevskij et al. and Kripjakevic et al. in 1963.[47] TiCoSb becomes a prospective p-type TE material owing to the semiconducting nature with VEC 18, chemical and thermal stability, mechanical strength, non-toxicity, cost effectiveness, and intriguing transport properties. The fascinating properties, contributing to the ZT, are strongly integrated with constituent atoms at specific Wyckoff positions.[48, 49] The Heusler alloys, X$_2$YZ type, are crystallized in the L2$_1$ structure, is schematically presented in the figure 1. However, the focus of interest as TE material, HH alloys, XYZ type, are crystallized in the C$_1$b structure.[31] XYZ type HH alloys may be considered as removing one of the X sites in the unit cell of Heusler alloy, having L2$_1$ structure, which consists of four fcc sublattices. The antisite defect or disorder in Heusler alloys may lead to a different structure.[31, 50] The inverse Heusler alloys, formed by replacement of constituent elements, and defect or disorder lead to unique properties owing to the drastic change in the electronic and thermal transport properties of the materials.[51-53] However, HH alloys are also crystalized in different inequivalent atomic orders.[31] The crystal structure of an HH alloy specifically depends on the Wyckoff positions and the constituent elements.[1, 2, 4, 7-9, 47] TiCoSb HH alloy with semimetallic-resistivity has been synthesized by Xia et.al., and the X-ray diffraction data have been indexed considering crystallization of the alloy as MgAgAs structure.[48] Sekimoto et al. have reported maximum ZT for TiCoSb, having a minute amount of impurity phases, at 921K and considered the crystal structure as MgAgAs.[47] Baker et al. have investigated the effect of pressure on the structural and TE properties of MgAgAs-type TiCoSb HH alloy, having Wyckoff positions of Ti, Co and Sb are 4a (0, 0, 0,), 4c ($\frac{1}{4},\frac{1}{4},\frac{1}{4}$) and 4b($\frac{1}{2},\frac{1}{2},\frac{1}{2}$), respectively, and no change in crystal structure owing to the pressure is reported.[1] However, theoretical analysis is also performed to understand the underlying mechanism to enhance the ZT of TiCoSb-based HH alloys. H. C. Kandpal et al. have analyzed the detailed electronic structure of the TiCoSb alloy, considering another positions of Ti, Co, and Sb, i.e., 4b ($\frac{1}{2},\frac{1}{2},\frac{1}{2}$), 4c ($\frac{1}{4},\frac{1}{4},\frac{1}{4}$), and 4a (0, 0, 0,).[44] A theoretical investigation for TE properties of thirty-six XYZ-type HH alloys, including TiCoSb are performed, considering X and Z form a rock salt structure and Wyckoff positions are 4b ($\frac{1}{2},\frac{1}{2},\frac{1}{2}$), 4c ($\frac{1}{4},\frac{1}{4},\frac{1}{4}$), and 4a (0, 0, 0,).[54] The effect of multi-site substitution, considering the atomic sites 4b ($\frac{1}{2},\frac{1}{2},\frac{1}{2}$), 4c ($\frac{1}{4},\frac{1}{4},\frac{1}{4}$) and 4a (0, 0, 0,), in TiCoSb is explored theoretically by Choudhary et al.[55] However, it is crucial to mention that there is only one theoretical investigation, best of our knowledge, considering two inequivalent positions of the atoms of the TiCoSb i.e., (i) 4a (0, 0, 0,), 4c ($\frac{1}{4},\frac{1}{4},\frac{1}{4}$) and 4b($\frac{1}{2},\frac{1}{2},\frac{1}{2}$), and (ii) 4c ($\frac{1}{4},\frac{1}{4},\frac{1}{4}$), 4a (0, 0, 0,) and 4d ($\frac{3}{4},\frac{3}{4},\frac{3}{4}$).[56] The theoretical and experimental studies of TiCoSb reveal that the atomic positions of the constituent atoms of TiCoSb may be different. The studies mentioned in this section are limited to identifying the atomic positions of the crystallized TiCoSb and corresponding electronic and TE properties. A systematic study regarding the atomic positions of a synthesized TiCoSb HH alloy is crucial to understand the TE behavior and enhancement of ZT.

There is a possibility of crystallization of HH alloy in different sets of Wyckoff positions and physical properties depend on the atomic positions in the crystal. Structural and TE properties of TiCoSb, being an HH alloy, are strongly corroborated with Wyckoff positions. To date, various Wyckoff positions are reported for TiCoSb HH alloy. However, an in-depth and systematic investigation, including structural optimization for possible atomic positions and their corresponding effect in electronic structure and TE properties of the TiCoSb alloy, is required. An attempt is taken to carry out a comparative analysis of four different sets of Wyckoff positions for TiCoSb HH material and find out the most probable structure of the synthesized TiCoSb compound. Rietveld refinement of the synthesized sample is performed, considering four sets of Wyckoff positions. Micro-structural characterization and in-depth structural characterization are performed employing energy dispersive X-ray spectroscopy (EDS) and transmission electron microscopy (TEM) of the synthesized sample. The sets of structural parameters, obtained after the Rietveld refinement are employed for the further theoretical calculations, employing the Quantum espresso package. The structural and electronic properties of the TiCoSb alloy are investigated for different Wyckoff positions using the Density Functional Theory (DFT) calculation. Additionally, the transport properties corresponding to the most probable structure are explored to delve into their TE behavior. In this endeavor, the detailed sample synthesis procedures as well as the experiments conducted for structural characterization are mentioned in section II. The optimized parameters, utilized to determine structural, electronic and TE transport properties by first principle calculation, are addressed in section III. Therefore, the results of experimental analysis and theoretical investigation are elaborately discussed and correlated in section IV. Finally, the conclusion, based on comprehensive analysis of experimental and theoretical data, is drawn in section V.

## II. Experimental details

The ingot of TiCoSb HH alloy was prepared by arc melting and soaking of the constituent elements, Titanium (Ti; 99.998% metal basis; Alfa Aeser, UK), Cobalt (Co) and Antimony (Sb) (each of 99.99% metal basis; Alfa Aesar, UK).[57] 2% excess Sb was deliberately incorporated in TiCoSb stoichiometry during the arc melting to mitigate the loss of Sb due to evaporation. The arc melting process under an argon atmosphere was performed for several times to achieve homogeneity. The arc melted ingot was sealed at vacuum in a quartz ampoule under a pressure of $10^{-3}$ mbar for soaking. The evacuated quartz ampoule was annealed at 1173 K for five days and then it was cooled down to room temperature at a rate of 10K/hour. A portion of the ingot was crushed into fine powder to carry out the X-ray diffraction (XRD). XRD was performed at room temperature in the range of 20°≤2θ≤100°, using a diffractometer model: X'Pert Powder, PANalytical, with Cu-K$_\alpha$ radiation of 1.5406 Å wavelength. In-depth structural characterization was performed, employing a mix-phase Rietveld refinement technique using FullProf software.[58] In order to explore the most probable structure of the synthesized sample, the refinement technique was carried out, considering three interpenetrating face centred cubic structure with four sets of different possible Wyckoff positions for the synthesized TiCoSb HH alloy.

Micro-structural characterizations were conducted employing a scanning electron microscope (SEM) and a transmission electron microscope (TEM) analysis. SEM graphs along with compositional information, using energy dispersive X-ray (EDX) measurements, were performed utilizing ZEISS EVO-MA 10 microscope. The TEM images and the Selected Area Diffraction (SAED) patterns are obtained using Titan G$^2$ 30 600, operated at 300 kV instrument. Fine powder sample is drop casted in a carbon supported TEM grid and dried overnight for TEM measurements. TEM images were analysed to reveal the crystallographic properties of the synthesized TiCoSb, utilizing ImageJ software.[59]

### III. Computational Details

The optimized lattice configuration and electronic structure of the synthesized alloy are investigated by implementing the first principle calculation using the Full Potential Linearized Augmented Plane Wave method (FP-LAPW) in Quantum Espresso package.[60, 61] FP-LAPW method is one of the most accurate implementation of Density Functional Theory (DFT).[62] The Generalized Gradient Approximation (GGA) technique with Perdew-Burke-Ernzerhof (PBE) potential is utilized as exchange-correlation functional to achieve the appreciable results with accuracy in self-consistency field (SCF) calculation.[63, 64] In order to find out the most probable structure of the synthesized TiCoSb HH alloy, structural energy is optimized by implementing the SCF calculation method on a conventional unit cell.[56, 65, 66] The structural optimization is performed by varying the lattice parameter within the range of approximately ± 5% of the value obtained from the Rietveld refinement of the XRD data of the synthesized TiCoSb alloy. The SCF calculations using the above-mentioned lattice parameters are computed with the kinetic energy cut-off of 65 Ry for the expansion of the plane waves in basis set. An optimized 8×8×8 Monkhorst pack k-grid in a conventional unit cell is used for Brillouin zone integration. However, electronic band structure calculation is performed for a dense k-mesh of 16×16×16. The energy convergence value ~10$^{-6}$ Ry is considered for the Brillouin zone integration and electronic band structure calculations. The partial density of states (PDOS) and total density of states (DOS) are estimated to reveal the electronic structure and position of the Fermi surfaces for the four probable HH structure with different sets of Wyckoff positions of TiCoSb alloy. Further, an attempt is taken to determine the transport properties, using the data file generated during DFT calculation as the input in BoltzTraP2.[67] Temperature dependent Seebeck coefficient (S), electrical conductivity (σ) and electronic thermal conductivity (κ$_e$) are estimated, theoretically using BoltzTraP2, as a function of chemical potential (μ).

### IV. Results and Discussion

#### a. Structural characterization: X-Ray Diffraction (XRD)

The XRD pattern of the synthesized TiCoSb sample is presented in figure 2. All the diffraction peaks are indexed with the space group F-43m. The Williamson-Hall method is used to estimate average grain size (~154 nm) and compressive lattice strain ($-5.09\times10^{-4}$). The necessary details of the calculations for estimating the grain size and lattice strain are provided in the supplemental information section I (Figure S1).

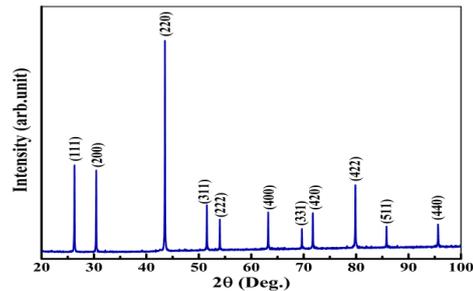

Figure 2: XRD pattern of synthesized TiCoSb alloy at room temperature. The lattice planes are identified and indexed as (h k l).

Structural characterization, using the XRD data, is performed employing Rietveld refinement, considering probable structures. In order to investigate the appropriate atomic orientation, four types of atomic positions (Table 1) are taken into consideration during the Rietveld refinement. The structures of the conventional unit cell of TiCoSb HH material with different atomic orientations are intensity of the peaks of type I structure does not match properly with the experimental data throughout the measured 2θ range, $10^0 \leq 2\theta < 100^0$. The experimentally observed intensity of most intense peak is perfectly fitted by the theoretical modelling, considering type II (Table1) structure of TiCoSb alloy. However, a discernible difference in the intensity as well as in the peak shape at higher angle of the diffraction pattern is observed. A good fitting quality after the refinements of the XRD data, using Wyckoff positions of type III and IV of TiCoSb alloy in the table1, is observed. The statistical parameters, which reflect the fitting quality, are presented in the table 2. A minute observation of the fitted XRD data, using theoretical modelling, and depicted in figure 3. The graphs, obtained after the refinements, are presented in the figure 4. The Rietveld refinement reveals that the calculated statistical parameters reveal that Wyckoff positions of type IV structure maximum aligns with the structure of synthesized TiCoSb sample.

Table 1: Four sets of Wyckoff positions for the probable structures of TiCoSb half-Heusler alloys. The Wyckoff position 4a, Octahedral position 4b and one of the tetrahedral positions 4c or 4d, are occupied in the crystal structure

| Structure Type | 4a (0,0,0) | 4b (½, ½, ½) | 4c (¼, ¼, ¼) | 4d (¾, ¾, ¾) |
|---|---|---|---|---|
| I | Ti | Sb | Co | - |
| II | Co | Ti | - | Sb |
| III | Ti | Co | Sb | - |
| IV | Sb | Ti | Co | - |

Table 2: Refinement parameters of four types of phases considered for TiCoSb, HH alloy.

| Structure | $R_p$ | $R_{wp}$ | $R_{exp}$ | Chi$^2$ |
|---|---|---|---|---|
| Type I | 12.2 | 20.9 | 5.35 | 15.3 |
| Type II | 13.9 | 22.3 | 5.36 | 17.3 |
| Type III | 9.11 | 16.1 | 4.97 | 10.5 |
| **Type IV** | **8.17** | **11.4** | **3.79** | **9.07** |

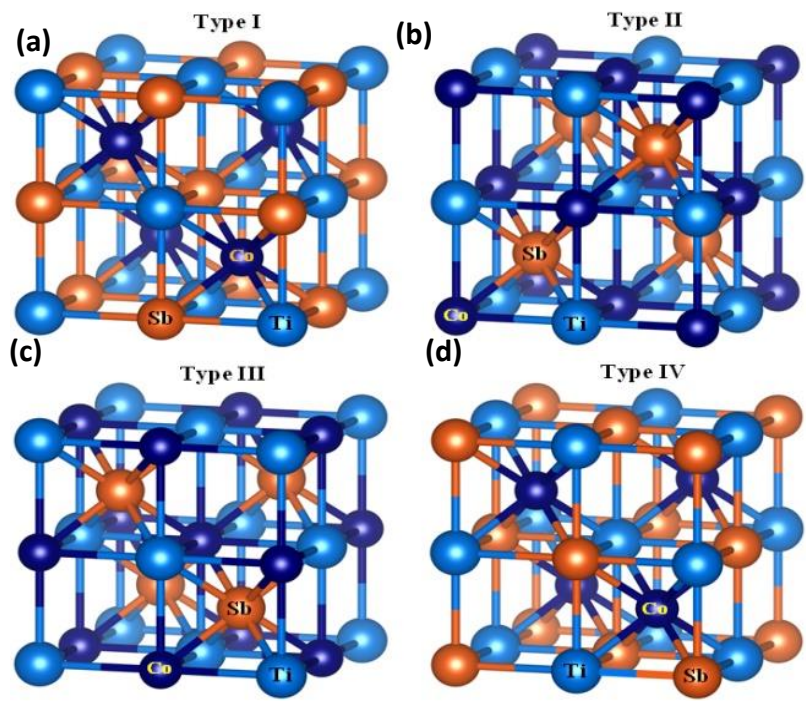

Figure 3: Four probable structures of conventional unit cell of TiCoSb half-Heusler alloy (a) Type I, (b) Type II, (c) Type III and (d) Type IV atomic orientation.

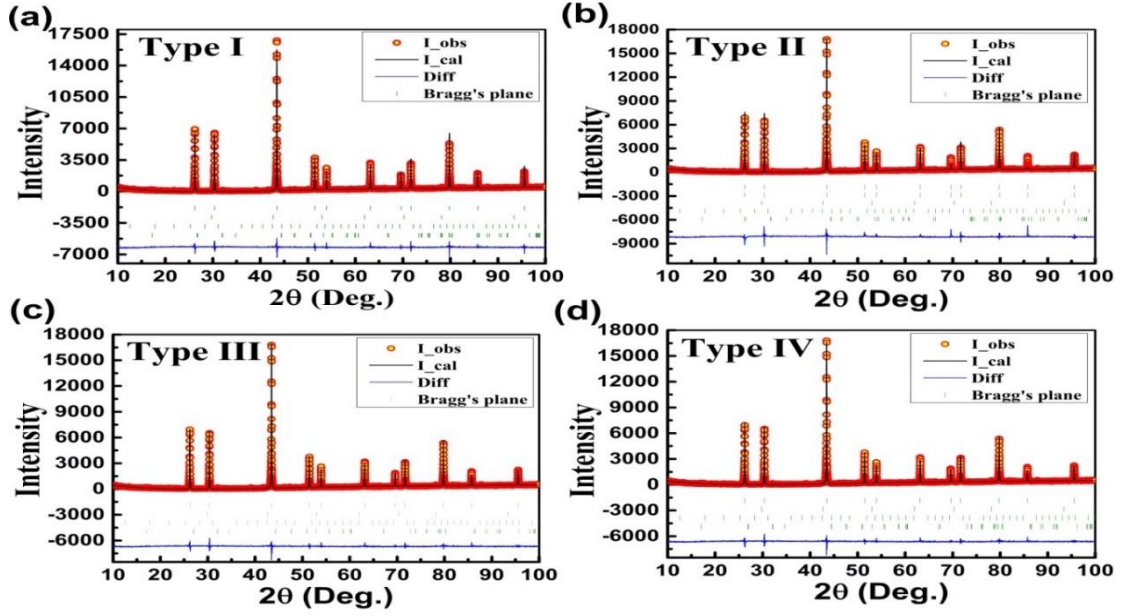

Figure 4: Mix-phase Rietveld refinement of (a) Type I (b) Type II (c) Type III and (d) Type IV. The red bubbled line depicts the experimental data, black and blue solid lines represent the theoretical data model and difference between the recorded and theoretical data respectively. The green vertical lines show the Bragg's plane of reflection.

A minute amount of TiCo, $Ti_2Co$ and CoSb embedded phases are also identified in the TiCoSb matrix after the mix-phase Rietveld refinement and data are provided in the section II of supplemental information.[47] An attempt is taken to estimate the lattice thermal conductivity ($\kappa_L$) by calculating the Debye temperature ($\theta_D$) utilizing the Debye-Waller factor ($B_{iso}$), obtained after the Rietveld refinement. The $B_{iso}$ and $\theta_D$ are related by the following equation,[68]

$$B_{iso} = \left(\frac{6h^2}{Mk_B\theta_D}\right)\left\{\frac{1}{4} + \left(\frac{T}{\theta_D}\right)^2 \int_0^{\theta_D/T} \frac{x}{e^x - 1} dx\right\} \quad (1)$$

where h, M, $k_B$ and T are Planck's constant, mass, Boltzmann constant and absolute temperature respectively. Finally, $\kappa_L$ is determined using the $\theta_D$ by employing the Slack equation,[69-71]

$$\kappa_L = \Lambda \frac{\bar{M}\theta_D^3 \delta}{\gamma^2 n^{2/3} T} \quad (2)$$

where $\Lambda = \frac{2.43 \times 10^{-6}}{1-(0.514/\gamma)+(0.228/\gamma^2)}$, a coefficient depends on the Grüneisen parameter ($\gamma=2.13$[56]). The other parameters $\bar{M}$, $\delta$, n, and T in equation (2) refer to the average atomic mass in amu, cubic root of the average volume per atom, number of atoms in a unit cell and absolute temperature, respectively. The estimated values of Debye temperatures and corresponding $\kappa_L$ are presented in table 3.

### b. Micro-structural characterization: EDX and TEM

The SEM images at different positions of the sample are provided in the supplemental information figure S2. Elements, present in the synthesized sample, are observed in the EDX graph, which is provided in the figure 5. The atomic weightage of constituent elements is determined in EDX measurement which confirms the 1:1:1 stoichiometry of the synthesized TiCoSb alloy (fig. 5). A negligible amount of gold (Au) is detected due to the thin layer of Au coating.

Table 3: The values of Debye temperature ($\theta_D$) and lattice thermal conductivity ($\kappa_L$) estimated from Debye-Waller factor ($B_{iso}$)

| Structure type | $B_{iso}$ (Å²) | $\theta_D$ (K) | $\kappa_L$ (W/mK) |
|---|---|---|---|
| I | 0.37765 | 445.5 | 7.15978 |
| II | 0.17386 | 618.4 | 19.14695 |
| III | 0.63939 | 362.8 | 3.86709 |
| **IV** | **0.61103** | **369.1** | **4.07208** |

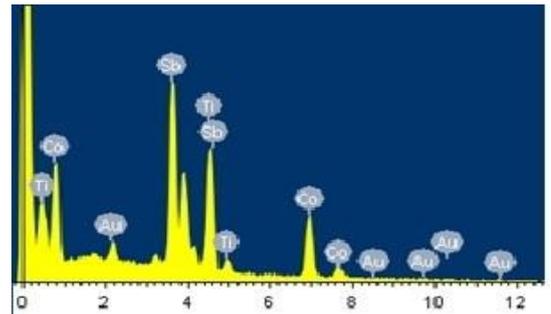

Figure 5: Graphical representation of EDX data of synthesized TiCoSb HH alloy

In order to explore and analyse the microstructure, TEM measurement of the synthesized sample is performed. TEM images of the sample at different position are provided in the supplemental information figure S3. However, a typical TEM image and selected area electron diffraction (SAED) pattern are shown in the figure 6. The analysis is accomplished by utilizing ImageJ software.[59] A continuous Debye rings in figure 6(a), the SAED pattern, of the synthesized TiCoSb compound is observed. A Debye ring is captured whenever the Debye cone, a cone of diffraction formed with the Bragg angles corresponding to the lattice planes, impinges upon the microscopic detector.[72] However, there is a limitation that the formation of Debye cones occurs only when the reciprocal lattice points are on the surface of Ewald sphere. It indicates that the diffraction phenomena takes place only for those reciprocal lattice vectors which intercept the Ewald sphere.[72, 73] The Ewald sphere is a geometrical representation of all the possible direction of diffracted beams corresponding to the reflection planes. It is formed at the origin of the reciprocal space with a radius of $1/\lambda$ ($\lambda \rightarrow$ the incident wavelength).[73] Each bright spot of diffraction is observed whenever the reciprocal lattice point coincides with the Ewald sphere surface fulfilling the Bragg's law, defined as,[74]

$$2d \sin \theta = n\lambda \quad (3)$$

where d, θ, n and λ denote the inter-planar spacing, angle of diffraction, order of diffraction and wavelength of incident beam respectively. The multiple bright spots, from various grains of the diffraction pattern, are superimposed with the Debye rings (Fig. 6(a)), provide the evidence of polycrystalline nature of the synthesized material.[75] The radius of the Debye rings, related to the reciprocal lattice, is proportional to the d-spacing of lattice planes in real space.[72] The diameter of Debye rings are estimated employing the ImageJ software and followed by the radius calculation. The d-spacings are calculated using the radius of the rings in the SAED pattern and the corresponding family of lattice plane are obtained by comparing with d-spacing in JCPDS data. The family of lattice planes are recognised as (111), (200), (220), (311), (222), (400), (331), (422), (511) and (440). It is crucial to mention that the lattice planes obtained from the analysis of SAED pattern, allowed for a cubic structure, are corroborated with the planes identified in the XRD pattern of the synthesized samples. The estimated values of d-spacing and their corresponding (h k l) planes are presented in table 4. High resolution TEM (HRTEM) images are studied to delve into the detailed information regarding the morphology of the synthesized sample presented in (fig. 6(b)). The XRD data and TEM data confirm the polycrystalline cubic structure of the synthesized TiCoSb sample.[75, 76] Four regions R1, R2, R3 and R4 of the HRTEM image, indicated in figure 6(b), are selected to explore the microstructural information in detail. Inter-planer spacing of the regions are extracted, employing the ImageJ software. Overlapping planes in the selected areas are observed and indicated in the figure 6(b). Crucial to mention, lattice planes are in good agreement with that of the space group of TiCoSb HH alloy.[76]

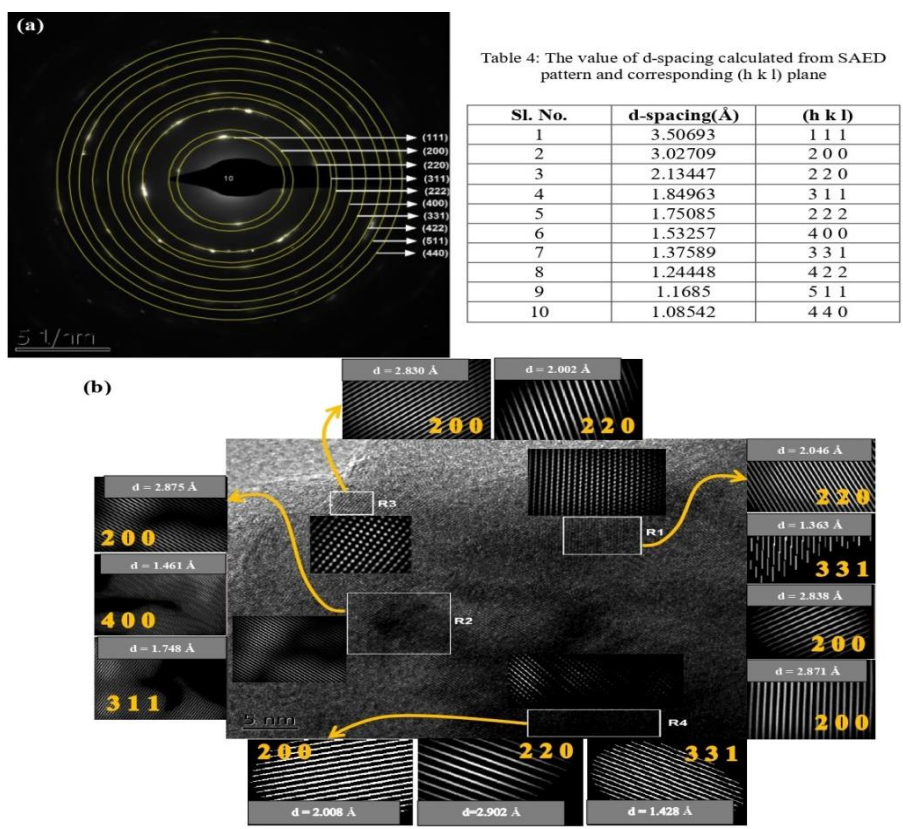

Figure 6: (a) SAED pattern of synthesized TiCoSb alloy showing the Debye rings, which is the evidence of crystalline nature of the sample (b) HRTEM image, where the four analyzed regions are indicated. The image of Fast Fourier Transform (FFT) and Inverse Fast Fourier Transform (IFFT) are also presented for each region. The identified (h k l) planes of all the regions are mentioned as well with their interplanar spacing.

Table 4: The value of d-spacing calculated from SAED pattern and corresponding (h k l) plane

| Sl. No. | d-spacing(Å) | (h k l) |
|---|---|---|
| 1 | 3.50693 | 1 1 1 |
| 2 | 3.02709 | 2 0 0 |
| 3 | 2.13447 | 2 2 0 |
| 4 | 1.84963 | 3 1 1 |
| 5 | 1.75085 | 2 2 2 |
| 6 | 1.53257 | 4 0 0 |
| 7 | 1.37589 | 3 3 1 |
| 8 | 1.24448 | 4 2 2 |
| 9 | 1.1685 | 5 1 1 |
| 10 | 1.08542 | 4 4 0 |

## c. Theoretical Investigations

### Structural optimization

An attempt is taken to explore the crystallized-structure theoretically and to support the experimental investigation, structural energy optimization is performed, implementing the DFT calculation using Quantum Espresso software.[60, 61] In order to perform the theoretical investigations, the energy optimization of the considered probable HH structures, given in table 1, employing the SCF calculations are carried out. The cell parameters of TiCoSb (Table 5) as obtained after

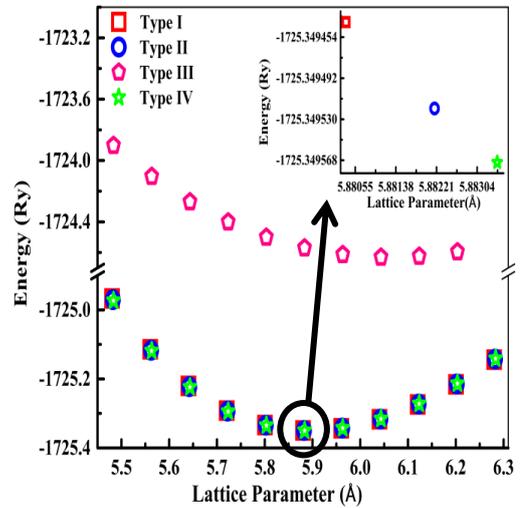

Figure 7: Optimized energy vs. lattice parameter curve, fitted with Murnaghan equation of state, for Type I, Type II, Type III and Type IV structures are represented by red square, blue circle, pink pentagon and green asterisk respectively. The inset depicts the energy of Type I, Type II and Type IV at equilibrium volume.

the Rietveld refinement for the probable Wyckoff positions (Table1) of the TiCoSb HH alloys are varied within ± 5% and corresponding SCF calculations are performed. The crystal structure of

Table 5: The estimated lattice parameters of TiCoSb alloy for the four types of atomic positions obtained after Rietveld refinement (a) and structural optimization ($a_0$), structural energy, bulk modulus (B) and pressure derivative of B, using the data obtained from structural optimization

| Structure type | Refined a (Å) | Optimized $a_0$ (Å) | Structural energy (eV) | B (GPa) | $\frac{dB}{dP}$ |
|---|---|---|---|---|---|
| I | 5.88036 | 5.89552 | -1725.34944 | 139.4 | 4.41 |
| II | 5.88217 | 5.89552 | -1725.34952 | 139.4 | 4.41 |
| III | 5.88341 | 6.06094 | -1724.63152 | 147.8 | 2.56 |
| IV | 5.88345 | 5.89548 | -1725.34957 | 139.4 | 4.41 |

TiCoSb HH alloy is cubic in nature.[31, 47, 65] Crystallization energy, obtained after the SCF calculations, of the considered lattice parameters for the probable structure is plotted (Fig. 7) as a function of lattice parameters and the curve is fitted by the Murnaghan equation of state (EOS) in Quantum Espresso, defined as,[77]

$$E(V) = E(V_0) + \frac{B_0 V}{B'_0}\left[\frac{(V_0/V)^{B'_0}}{B'_0 - 1} + 1\right] - \frac{B_0 V_0}{B'_0 - 1} \quad (4)$$

where, E is the energy at a particular volume, $V_0$ is the volume at equilibrium, $B_0$ and $B'_0$ are the bulk modulus and pressure derivative of bulk modulus, respectively. Figure 7 indicates the change in structural energy of the probable structures with the variation of the lattice parameter. The optimized structural energy and their corresponding cell parameter are mentioned in table 5. The equilibrium lattice parameter, considering the theoretical structural optimization, is higher than that of the Rietveld refinement. The maximum optimized structural energy is obtained for type III structure. However, the structural energy of type I, type II and type IV is comparable and careful investigation (magnification of the equilibrium points, inset of fig. 7) reveals that type IV structure of the TiCoSb HH alloy has the minimum energy with a lattice parameter of a=5.88345Å.[3, 44, 47, 56, 65, 78] The Bulk modulus (B) and pressure derivative of Bulk modulus (dB/dP), calculated in Murnaghan EOS, are comparable (B=139.4 GPa) for all the considered structures except type III. The estimated B values are nicely corroborated with the previous reports.[44, 56, 65, 66]

**Electronic structure**

The electronic bands of the probable Wyckoff positions of the synthesized material are explored by evaluating DOS and band structure calculation. Figure 8(a-d) represent the partial DOS (PDOS), total DOS of the considered structures for TiCoSb HH alloys and the Fermi energy level ($E_f$) is denoted at 0 eV. Crucial to note that maximum contribution to the DOSs is obtained from the outer most orbital of each atom i.e., Ti-3d, Co-3d and Sb-5p. Figure 8 indicates that there is no trace of DOS near the $E_f$ for probable structures, type I (a) and type IV (d). However, presence of the electronic states at the $E_f$ is observed for the type II (fig. 8 (b)) and type III

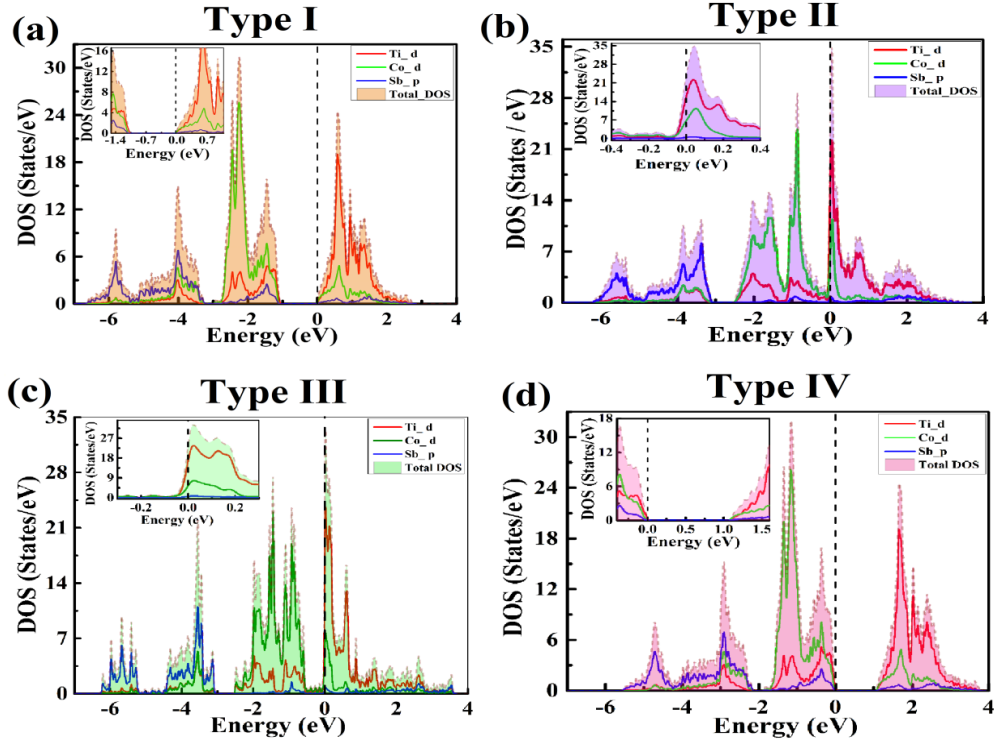

Figure 8: Partial Density of states (PDOS) and total Density of States (DOS) of different structures of TiCoSb HH alloy (a) Type I, (b) Type II, (c) Type III and (d) Type IV. The insets of (a), (b), (c) and (d) depict the DOS at $E_f$ level

(fig. 8 (c)). The analysis of DOS graphs, estimated for the considered structures, clearly reveal that the type I and IV structures are semiconducting in nature, while the structures of type II and III are metallic. However, minute observation of figure 8 reveals that $E_f$ is shifted towards the CB in type I, while in type IV it is towards the VB. The positions of $E_f$ in the DOS graphs indicate TiCoSb HH alloy with Wyckoff positions corresponding to type-I is n-type and the type-IV is p-type.[79]

The structural investigation through XRD analysis employing Rietveld refinement reveals that the statistical parameters obtained after refinement is best for type IV. Eye estimation of the refinement also suggest that XRD data is best fitted by employing the Wyckoff positions of the type-IV structure. It is crucial to note that first principle calculation employing Quantum Espresso software also indicates lowest structural energy, estimated using the lattice parameters, obtained after refinement, is minimum for the probable structure, type-IV. The density of states and electronic structure, estimated by first principal calculations, strongly supports that type-IV structure is the only p-type semiconductor, which aligns with the established and reported physical properties of the TiCoSb HH alloy.[56] The further theoretical calculations, to reveal the electronic and TE properties of TiCoSb, are performed, considering the Wyckoff positions of type-IV. $\kappa_L$ is estimated by employing the equation (2) (Table 3). It is crucial to mention that a very low $\kappa_L$ is obtained for the synthesized material with the most favourable, type-IV, crystal structure of TiCoSb. The estimated value of the $\kappa_L$ is 4.07208 W/mK. A comparative study of various previously reported data and estimated value of $\kappa_L$ are graphically represented in figure 9.[3, 49, 56, 78, 80, 81] It depicts that our estimated $\kappa_L$ is quite lower than that of the reported theoretical and experimental investigations.

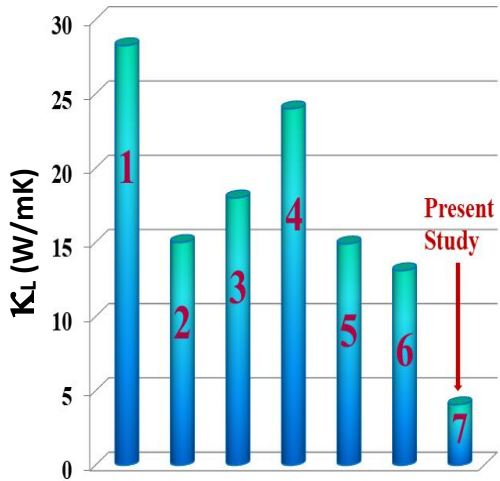

Figure 9: Pictorial representation of a comparative study for $\kappa_L$ between previously reported data (bars 1, 2, 3, 4, 5, and 6)[56, 80, 3, 49, 78, 81] and the present TiCoSb sample study.

In order to understand the contribution of atoms in formation of DOS of TiCoSb, PDOS for type IV is studied (fig. 8 (d)). The Co atom contributes the maximum to the electronic states, whereas Sb atom has the minimum contribution to the total density of states. It shows that more number of Co atomic states is occupied than the other. The figure 8 (d) depicts that Ti-3d and Co-3d have maximum contribution to the total DOS of the TiCoSb alloy. In order to elucidate the PDOS the valence band may be considered as two regions, lower valence region (-6 eV to -2 eV) and upper valence region (-2 ev to 0 eV). In the lower valence band region Ti-3d and Co-3d have almost equal contribution but Sb-5p has slightly higher contribution. The upper valence region has strong contribution of Co-3d state, while Ti-3d and Sb-5p show considerable contribution to this region. The conduction band region (1 eV to 4 eV) shows sharp peaks of 3d state of Ti atom. It has a considerable amount of Co-3d state but the contribution of Sb-5p state is negligible. The analysis of PDOS reveals that the electronic states in valence band region is due to the contribution of Co atomic state and conduction band region is mainly due to the contribution of 3d state of Ti atom in TiCoSb HH alloy.

Figure 10(a-c) represents the electronic band structure for the TiCoSb HH alloy, having Wyckoff positions of type I, III and IV (Table1) respectively. The electronic band structure is plotted along the high symmetry directions W-L-Γ-X-W-K. The $E_f$ is indicated by a dashed line at 0 eV, located near the VB for type-IV in figure 10(c). The band structure of type-IV (fig. 10(c)) reveals that the valence band maxima (VBM) and conduction band minima (CBM) are located at Γ

point, indicating direct band gap energy of 1.09 eV.[54-56, 65] In addition to this, the effective mass relative effective mass as 1.48 for TiCoSb HH material.[56]

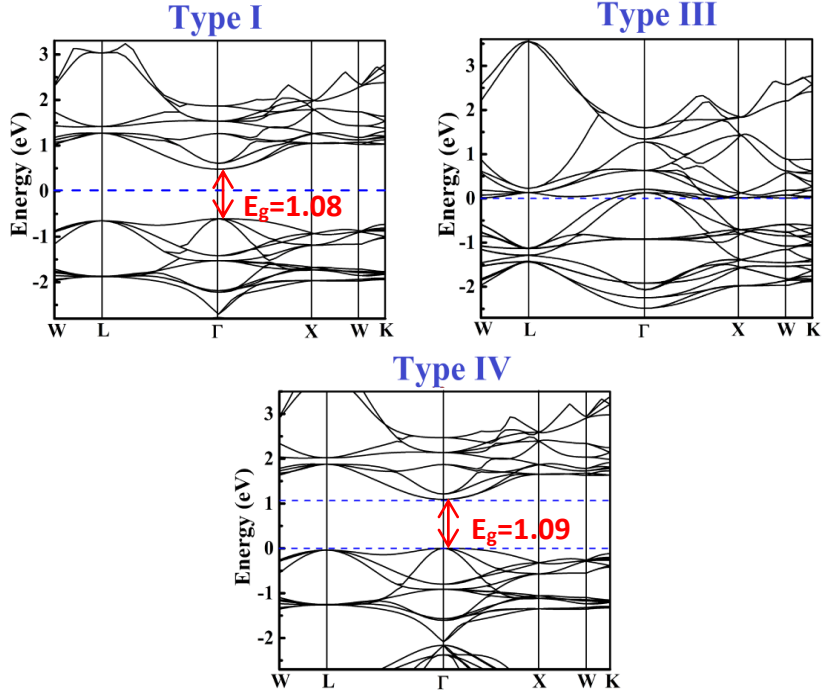

Figure 10: Band structure of TiCoSb HH alloy for, (a) Type I, (b) Type III and (c) Type IV structures

($m^*$) of TiCoSb is also determined by fitting the energy points in the lowest edge of CB. Semiconducting behaviour in a material is determined by the electrons near to the most narrow band gap.[82] The relation between energy (E) and wave vector (**k**) near the minimum band gap may be written as, considering the parabolic approximation,[83]

$$E(k) = Ak^2 + Bk + C \quad (5)$$

i.e.,
$$\frac{d^2E}{dk^2} = 2A \quad (6)$$

where, A is the coefficient of $k^2$, indicate the curvature of the parabola. B is the linear term and C represents energy at band edge (i.e., CBM). The free electron mass is replaced by the effective mass ($m^*$) for a crystal lattice [82] as,

$$m^* = \frac{\hbar^2}{2A} \quad (7)$$

Here $\hbar = h/2\pi$; where h is the Plank's constant. Equation 7 is employed to estimate the $m^*$ and the estimated value of the relative effective mass ($m^*/m_e$) is 1.75. Joshi et al. have estimated the

## Thermoelectric Properties

Experimental and theoretical investigations indicate that the synthesized TiCoSb HH alloy is crystallized as the Wyckoff position of type IV structure. In order to understand the thermoelectric behaviour of TiCoSb, theoretically, for the type IV structure, an effort is given to determine the thermoelectric transport properties. In this attempt Seebeck coefficient (S), electrical conductivity (σ) and electronic thermal conductivity ($\kappa_e$) are studied by employing BoltzTraP2 package. The theoretical calculations are performed by the Boltzmann semi-classical transport equation, employing Fourier interpolation, utilizing rigid band approximation.[67, 84] However, the band structure of a material is considered invariable under the influence of temperature, but the chemical potential of that material changes.[84] The output of the first principle calculation, conducted using Quantum Espresso, yields the input dataset for the subsequent analysis of thermoelectric parameters in BoltztraP2. BoltztraP2 computes the electrical conductivity and electronic thermal conductivity as a ratio with relaxation time (τ), denoted as σ/τ and κ$_e$/τ. The transport properties are plotted in figure 11 as a function of μ for the temperature range 300K-1200K, and the E$_f$ is denoted with the dashed line at 0.00 eV. Figure 11 (a) depicts the S(μ) for the temperature range 300K to 1200K and the positive trend of Seebeck coefficient represents the involvement of p-type charge carriers (holes) below the E$_f$ level, while material at room temperature, representing at figure 11(a), exhibits an intrinsic semiconducting behaviour.[56, 85] Subsequently, the inset of figure 11(a) depicts the change in S at E$_F$ level with increasing temperature. It reveals that the maximum S ~376 μV/K is achieved at 500 K at μ=E$_f$ level and then gradually decreases with temperature. The typical behaviour of S at high temperature signifies the enhancement of the prominent negative trend indicates the strong influence of n-type charge carriers (electrons) for the region above the E$_f$ level. The increase in S(μ) is related to the decrement in the charge carriers.[65] The highest S is achieved, when μ equal to the energy of band edge.[56, 65] However, S drops to 0.0μV/K at μ= E$_f$. The high value of S for a specific μ, makes the TiCoSb HH alloy as a potential TE material. The typical variation of S(μ) at 300 K depicts a region of S=0 μV/K for TiCoSb, suggests no free electron or hole near the E$_F$ at room temperature.[56] S(μ) of TiCoSb HH thermally excited charge carriers of TiCoSb (electrons for μ<E$_F$ and holes for μ>E$_F$).[56] The maximum value of S ~ 675μV/K at 800K, theoretically estimated, is reported.[56] However, Sekimoto et al. have experimentally obtained the highest S~150μV/K at 350K[47] and ~360μV/K at 425K[86] for TiCoSb alloy, synthesized by solid-state reaction and SPS method respectively.

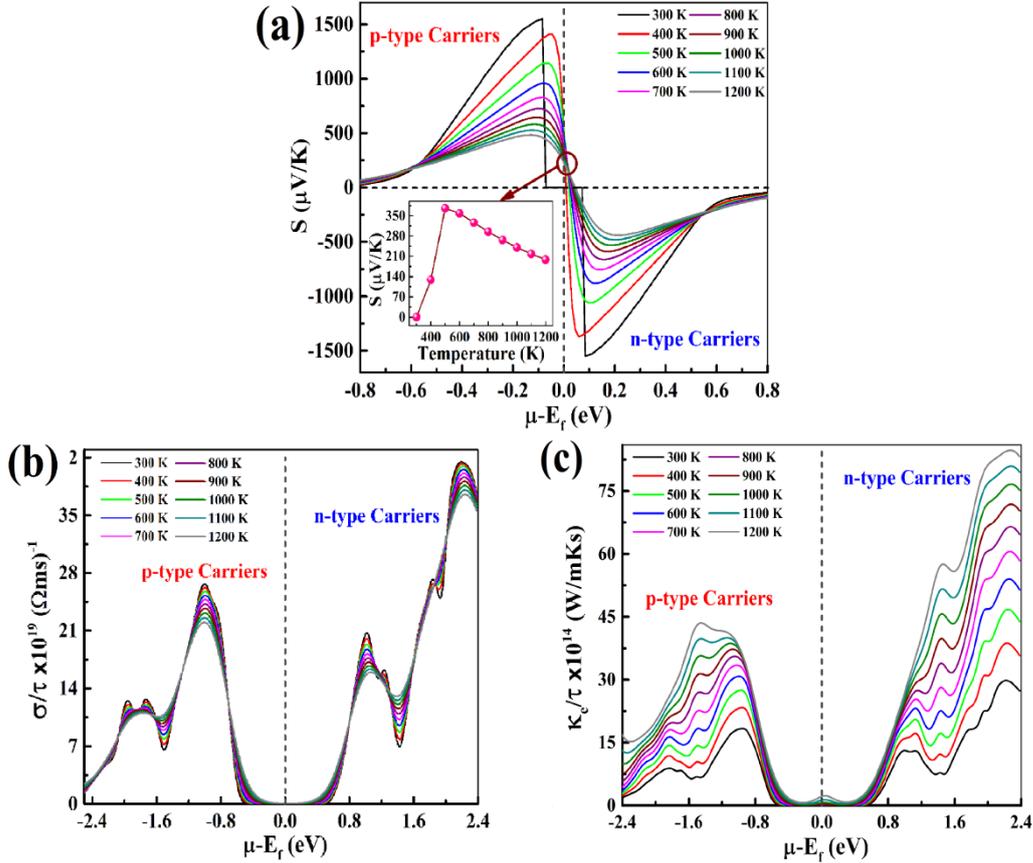

Figure 11: The variation of thermoelectric parameters with chemical potential (μ) at different temperatures. (a) Seebeck coefficient (S), inset depicts the change in S as a function of temperature at E$_f$. (b) electrical conductivity (σ/τ, τ=relaxation time) and (c) electronic thermal conductivity (κ$_e$/τ, τ=relaxation time).

Figure 11 (b) and (c) represents the temperature dependent variation of σ/τ and κ$_e$/τ as a function of μ. The increase in scattering rate of carriers attributes in reduction of the mean free path of charge carriers at very high temperature, leading to the slight decrease in σ/τ and limits the rate of increase of $\kappa_e/\tau$.[56] An attempt is taken to determine the electron relaxation time (τ) to study the behaviour of TE parameters with temperature. The τ depends upon various scattering mechanisms like electron impurity, electron-electron and electron-phonon scattering.[87, 88] As the ab-initio band structure calculation does not involve the DOS effective mass in evaluation of chemical potential, it is difficult to carry out a quantitative study of τ, dependent on both the temperature and μ. The τ is estimated for the specific value of μ, at which the room temperature S reaches at maximum. The value of τ is evaluated using the Heisenberg uncertainty principle and Drude model,[87]

$$\tau = A \frac{qSh}{k_B^2 T} \quad (8)$$

where, A is a dimensionless constant factor, depends on the characteristic of the material. The value of *A* is 1 and 0.1 for metal and semiconductor, respectively. The other parameters, i.e., q, S, h, $k_B$ and T are the charge of the carriers, Seebeck coefficient, Planck's constant, Boltzmann constant and absolute temperature, respectively. The value of τ for TiCoSb semiconducting alloy may be estimated by modifying the equation (8) as,[87, 89]

$$\tau = 0.1 \frac{eSh}{k_B^2 T} \quad (9)$$

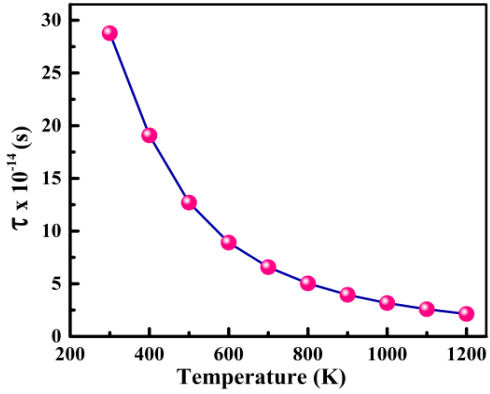

Figure 12: Temperature dependent relaxation time (τ) at the certain value of μ at which the room temperature S attains its highest value

The thermal variation of τ is presented in figure 12. Figure 13(a) represents the S (T). In addition to this, the change in σ (Fig. 13(b)), $\kappa_e$ (Supplemental Information fig.S4) and PF (fig. 13(c)) are calculated as a function of temperature. The PF (=S²σ) (fig. 13(c)) shows a significant increase from 500K onwards.

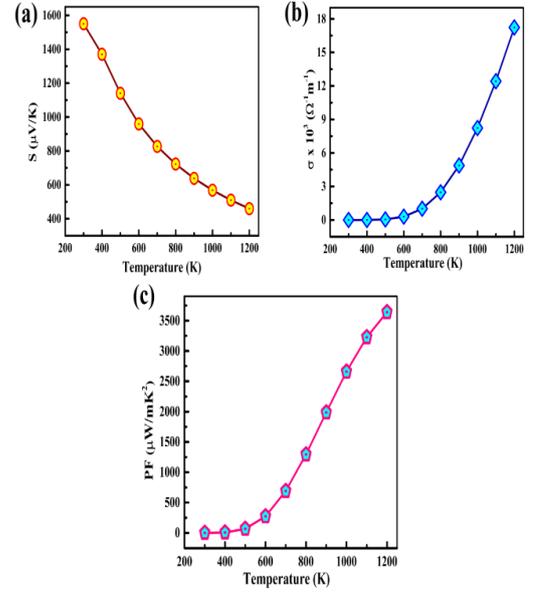

Figure 13: Variation in (a) Seebeck coefficient (S), (b) electrical conductivity (σ) and (c) power factor (PF) with temperature at a certain chemical potential for Type-IV structure of TiCoSb.

## V. Conclusion

The study reveals the crystal structure of the synthesized TiCoSb HH alloy among the four probable sets of Wyckoff positions for a HH alloys and estimates the TE parameters along with electronic properties for the synthesized TiCoSb alloy. A comparative study of different structural model with inequivalent atomic positions of synthesized TiCoSb HH alloy is carried out to explore the most probable crystal and electronic structure. The transport properties corresponding to that structural model is rigorously analysed. The structural characterization, employing Rietveld refinement confirms that the synthesized TiCoSb HH alloy is crystallized with F-43m space group. The microstructural characterizations, employing SEM and TEM, reveal the stoichiometry and Bragg's reflection planes of the synthesized polycrystalline lattice structure of TiCoSb HH alloy, corroborating the refinement results of the XRD data. The statistical parameters and fitting quality indicate that the Wyckoff positions of the synthesized TiCoSb alloy are Sb: 4a (0, 0, 0), Ti: 4b (½, ½, ½), Co: 4c (¼, ¼, ¼). Further, the optimization energy and lattice parameters of the crystal structure, DOS, and electronic structure, estimated theoretically, also support the experimental observations. The electronic structure calculation confirms that the synthesized sample is p-type with a band gap of 1.09 eV. It is crucial to mention that the $\kappa_L$ of the synthesized TiCoSb sample is minimum, compared to the reported values for the TiCoSb-based HH alloys.

The transport properties are estimated employing the first principle calculations and reveal that S at 300 K describes the intrinsic property of TiCoSb HH alloy due to the existence of no charge carriers at $\mu=E_f$. The temperature-dependent behaviour of $\sigma/\tau$ and $\kappa_e/\tau$ correlated with the semiconducting nature of the TiCoSb alloy. However, the thermal variation of $\tau$ is estimated to calculate the PF. PF of the synthesized material rapidly enhances after 500K.

TiCoSb, synthesized by solid state reaction followed by arc melting, is crystallized with F-43m space group, having Wyckoff positions Sb: 4a (0, 0, 0), Ti: 4b (½, ½, ½), Co: 4c (¼, ¼, ¼). Electronic structure and TE properties confirm that TiCoSb is a potential p-type semiconducting material with maximum efficiency at mid to high temperature.

## VI. Acknowledgement

The financial support of UGC-DAE-CSR Kalpakkam, India (Ref: CRS/2021-22/04/639, CRS/2022-23/04/893) is acknowledged in the form of research project grants. We gratefully acknowledge late Avijit Jana for his support in carrying out the experiments and engaging in valuable discussions. We remember his contributions with deep respect and sorrow.

# Supplemental material for "Favorable half-Heusler structure of synthesized TiCoSb alloy: a theoretical and experimental study"

Pallabi Sardar[1,2], Suman Mahakal[2], Soumyadipta Pal[4], Shamima Hussain[5], Vinayak B. Kamble[6], Pintu Singha[6], Diptasikha Das[1]* and Kartick Malik[2]*

[1]*Department of Physics, ADAMAS University, Kolkata-700 126, West Bengal, India*
[2]*Department of Physics, Vidyasagar Metropolitan College, Kolkata-700 006, West Bengal, India*
[4]*University of Engineering and Management, New Town, University Area, Plot No. III, B/5, New Town Rd, Action Area III, Newtown, Kolkata 700160, West Bengal, India*
[4]*Department of Physics, Institute of Engineering & Management, Management House, D-1, Sector-V, Saltlake Electronics Complex, Kolkata 700 091, West Bengal, India*
[5]*UGC-DAE Consortium for Scientific Research, Kalpakkam Node, Kokilamedu-603 104, Tamil Nadu, India*
[6]*Indian Institute of Science Education and Research (IISER) Trivandrum, Vithura, Trivandrum 695 551*

## Section I

- **Calculation of grain size and lattice strain using Scherrer equation and Williamson-Hall method:**

The X-ray diffraction (XRD) data are utilized to determine the average crystallite size (D) and lattice strain (ε) by employing the Scherrer equation and Williamson-Hall (W-H) method. The XRD peaks get broadened due to (i) crystalline size and lattice strain effects and (ii) instrumental parameters.[1,2] The instrumental broadening is measured by using the crystalline Silicon as a standard material for calibration of position.[2] The corrected broadening ($\beta_D^2$) is estimated by using the equation,

$$\beta_D^2 = \beta_{measured}^2 - \beta_{instrumental}^2 \quad (1)$$

where $\beta_{measured}$ and $\beta_{instrumental}$ are the measured broadening of the TiCoSb sample and instrumental broadening for standard Silicon respectively, at half-maximum intensity. The average grain size (D) of the crystal is related to the $\beta_D$ with the Scherrer equation,[2]

$$D = \frac{k\lambda}{\beta_D cos\theta} \quad (2)$$

where D, λ and θ are particle size in nm, wavelength of incident X-ray beam (1.54056Å for Cu $K_\alpha$ radiation) and peak position respectively and k (=0.9) is a constant.[1,2] However Scherrer equation involves the crystallite size only in calculation of peak broadening.

The Williamson-Hall method includes the microstructural effect (i.e., lattice strain), which develops in the crystal lattice due to crystal imperfections and distortions.[3,4] According to W-H method the total physical broadening of the Bragg peak ($\beta_{hkl}$) is additive in nature, i.e.,

$$\beta_{hkl} = \beta_D + \beta_{strain} \quad (3)$$

The strain-induced broadening ($\beta_{strain}$) is considered to be uniform throughout the crystallographic direction according to the uniform deformation model (UDM).[3] It is defined as,

$$\beta_{strain} = 4\varepsilon tan\theta \quad (4)$$

where, ε is the lattice strain. Hence, the equation (3) can be written as follows, using equation (2) and (4)

$$\beta_{hkl} = \frac{k\lambda}{D\cos\theta} + 4\varepsilon\tan\theta \quad (5)$$

On rearranging the equation (5), we get

$$\beta_{hkl}\cos\theta = \frac{k\lambda}{D} + 4\varepsilon\sin\theta \quad (6)$$

Now, it can be seen that the equation (6) corresponds to the equation of a straight line. Hence, the graphical representation of $4\sin\theta$ along x-axis versus $\beta_{hkl}\cos\theta$ along y-axis gives a straight line (Figure S1), which intercepts the y-axis at $k\lambda/D$ with a negative slope of ε. Since, the linear fitting of the estimated data points provides the intercept and slope of the graph, the average crystallite size of synthesized TiCoSb alloy is estimated by utilizing the values of k & λ and the compressive lattice strain is equal to the slope.

**Figure S2: Williamson-Hall plot for synthesized TiCoSb alloy**

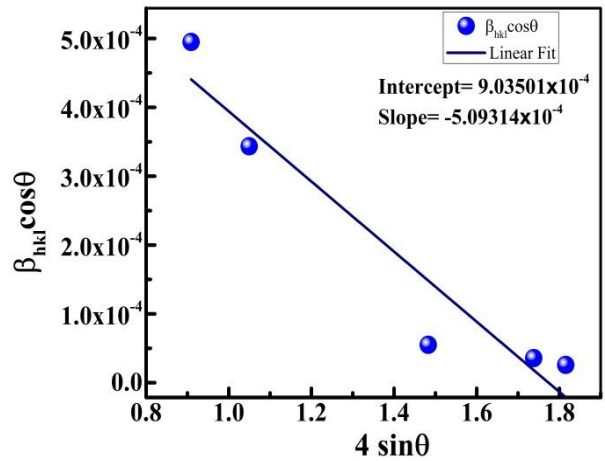

## Section II

- **The amount (wt%) of embedded phases present in the synthesized TiCoSb material, obtained from mix-phase Rietveld refinement**

| TiCoSb | TiCo | Ti$_2$Co | CoSb |
|---|---|---|---|
| 92.01±0.62 | 2.56±0.29 | 3.52±0.33 | 1.91±0.16 |

- **Figure S2: SEM images of synthesized TiCoSb half-Heusler alloy**

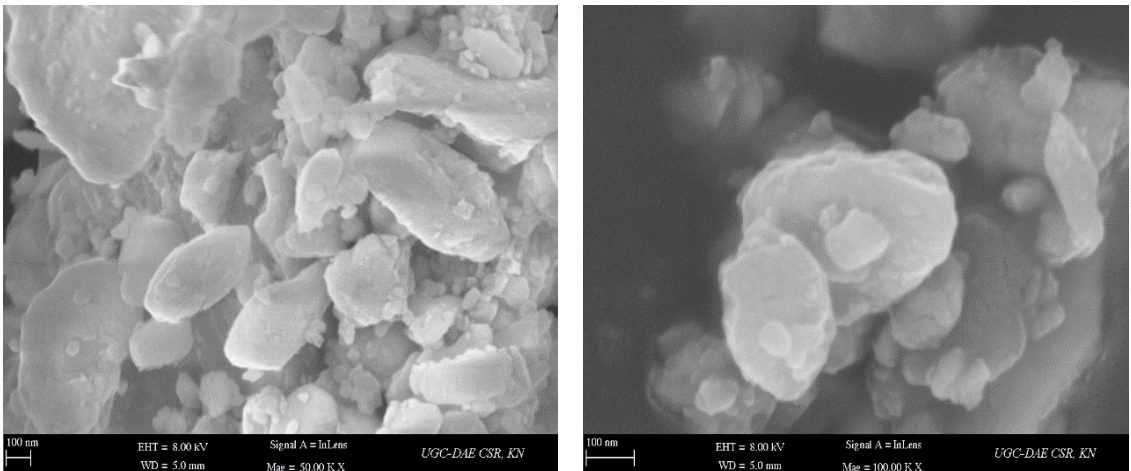

- **Figure S3: TEM images of synthesized TiCoSb half-Heusler alloy, captured at different positions**

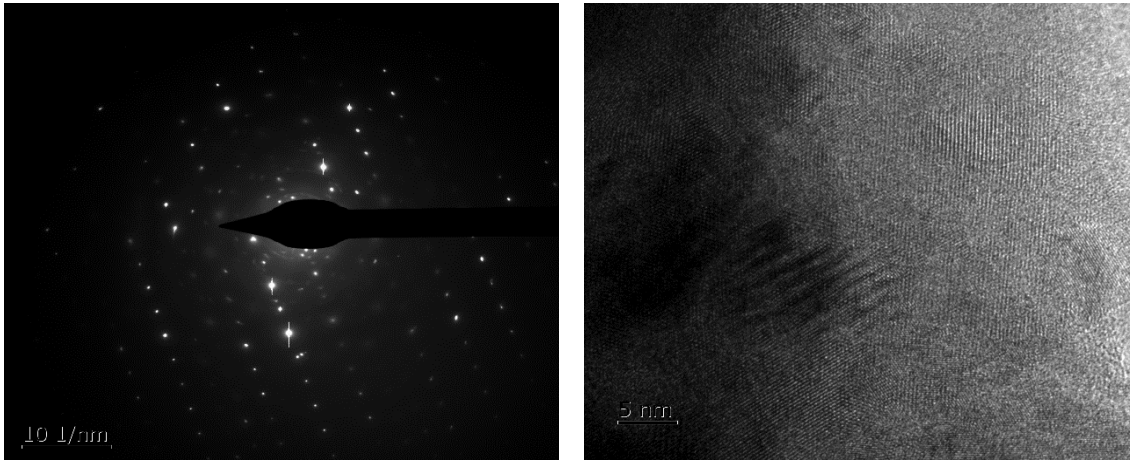

- **Figure S4: Electronic thermal conductivity ($\kappa_e$) of synthesized TiCoSb half-Heusler alloy as a function chemical potential, determined using BoltzTraP2**

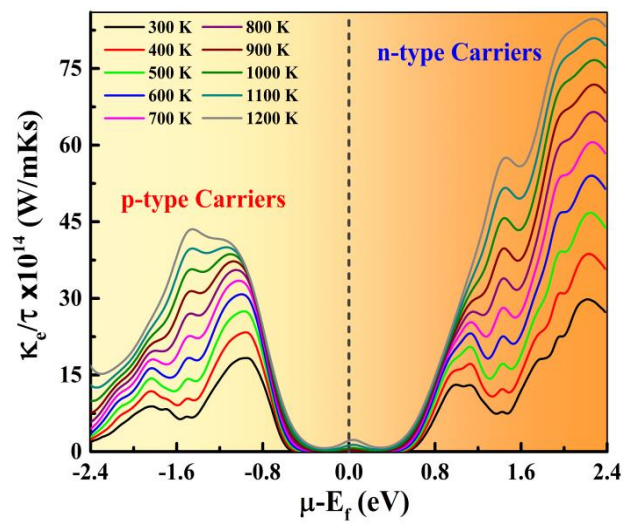

- **References**